\documentclass{svjour3}
\usepackage{graphicx}
\usepackage{natbib}
\begin{document}

\title{Measurement of the radial velocity of the Sun as a star by means of a reflecting solar system body}
\subtitle{The effect of the body rotation}
%\titlerunning{RV effect of a rotating reflecting body}
%\authorrunning{A.~F.~Lanza and P.~Molaro}
\titlerunning{Measuring the radial velocity of the Sun as a star by a reflecting body}
\author{A.~F.~Lanza \and P.~Molaro}
\institute{A.~F.~Lanza
\at INAF-Osservatorio Astrofisico di Catania, Via S.~Sofia, 78 - 95125 Catania, Italy\\
\email{nuccio.lanza@oact.inaf.it}
\and
P.~Molaro
\at INAF-Osservatorio Astronomico di Trieste, Via G. B. Tiepolo, 11 - 34143 Trieste, Italy\\
\email{molaro@oats.inaf.it}}

\maketitle
\begin{abstract}
Minor bodies of the solar system can be used to measure the spectrum of the Sun as a star by observing sunlight reflected by their surfaces. To perform an accurate measurement of the radial velocity of the Sun as a star  by this method, it is necessary to take into account the Doppler shifts introduced by the motion of the reflecting body. Here we discuss the effect of its rotation.  It gives a vanishing contribution only when the inclinations of the body rotation axis to the directions of the Sun and of the Earth observer are the same. When this is not the case, the perturbation of the radial velocity does not vanish and can reach up to $\sim 2.4$~m/s for an asteroid such as 2~Pallas that has an inclination of the spin axis to the plane of the ecliptic of $\sim 30^{\circ}$. We introduce a geometric model to compute the perturbation in the case of a uniformly reflecting body of spherical or triaxial ellipsoidal shape and provide general results to easily estimate the magnitude of the effect. 
\end{abstract}
\
\section{Introduction}
Obtaining a spectrum of the Sun integrated over its disc, i.e., directly comparable with stellar observations, is not an easy task. Generally, the spectrum reflected by a minor body of the solar system or by one of the Galileian satellites has been used as a proxy for the spectrum of the Sun as a star \citep[cf.][]{MolaroCenturion11}. An accurate measurement of the wavelength of a spectral line in a reference frame at rest with respect to the barycentre of the Sun, or an accurate measurement of the radial velocity (hereafter RV) of the Sun as a star, require a correction for the Doppler shift produced by the motion of the reflecting body \citep[cf.][]{MolaroMonai12,Molaroetal13}. The effect of the orbital motion with respect to the barycentre of the Sun and to the observer on the Earth can be corrected using the NASA Horizon ephemerides\footnote{http://ssd.jpl.nasa.gov/?horizons}, but the effect of the axial rotation of the body must also be  taken into account when a precision of the order of $0.1-1$ m/s is required. This is the case of the observations of the Sun as a star performed to understand the impact of solar convection and magnetic activity on its disc-integrated RV. These investigations are of fundamental importance  to understand similar effects in distant solar-like stars that are searched for Earth-like planets 
 \citep[e.g.,][]{Lanzaetal11,Dumusqueetal12}. 

This work is dedicated to a precise computation of the effect of the axial rotation of a reflecting body -- hereinafter indicated, for simplicity, as an asteroid -- on the solar radial velocity. 
For simplicity, we shall assume that the body has a spherical surface in Sect.~\ref{model}, while the case of a triaxial ellipsoidal shape will be considered in Sect.~\ref{model_elliptic}.  

We consider a body of uniform albedo. Patches with a different albedo on the surface of a rotating asteroid may produce distortions of the line profiles in the reflected spectrum that are modulated with the rotation period of the body. { The origin of the line distortions is the different level of the continuum in the spectrum reflected from an albedo inhomogeneity. For example, in the case of a dark spot, the local lower continuum level  produces a bump in the spectrum integrated over the disc of the asteroid as in the case of a spotted star \citep[cf. Fig.~1 in][]{VogtPenrod83}.} This  affects  the measured solar RV with a periodic perturbation having the same period of the rotation of the asteroid\footnote{{ In principle, this effect can be used to measure the rotation period of an asteroid if the amplitude of the modulation is comparable or larger than the intrinsic radial velocity variations of the Sun on typical rotational timescales, i.e.,  a few hours or days. Those intrinsic  variations are dominated by surface convection and have amplitudes of a few m/s \citep[cf.][]{Dumusqueetal11}.}}. It is possible to correct for this effect by fitting a sinusoid and its  harmonics to the RV time series with their fundamental period equal to the rotation period of the asteroid. The  coefficients of this Fourier series will be a slowly varying function of the viewing angle of the asteroid spin axis that implies that this method can be applied only for time series that are much shorter than the asteroid and the Earth orbital periods \citep{Haywoodetal15}. On the other hand, the effect that we investigate in the present work is not modulated with the rotation of the asteroid, but varies slowly with the relative position  of the body and of the Earth along their orbits. Therefore, it is not possible to correct for this effect by the simple method applicable in the case of the modulation arising from the albedo inhomogeneities. 
\section{Model}
\subsection{Reference frame}
We consider a  reference frame with the origin at the barycentre $T$ of the asteroid. 
Its spin axis  points in the direction $(\ell_{\rm S}, \beta_{\rm S})$, where $\ell_{\rm S}$ is the ecliptic longitude and $\beta_{\rm S}$ the ecliptic latitude of the projection of its North pole onto the celestial sphere as seen from  $T$. The ecliptic longitude of the Sun and of the (geocentric) observer, as seen from $T$ at the epoch of sunlight reflection\footnote{For simplicity, we can neglect the  delays due to the time taken by sunlight to travel from the Sun to the asteroid and be reflected to the observer because the intervening variations in the positions of the bodies are too small to significantly affect the angles entering in our model (see below).}, are indicated with $\ell_{\odot}$ and $\ell_{\oplus}$, respectively. The  inclinations of the asteroid spin axis to the direction of the Sun and to the observer lying on the plane of the ecliptic, $i_{\odot}$ and $i_{\oplus}$, respectively, are given by:
\begin{eqnarray}
\cos i_{\odot} = \cos (\ell_{\odot} - \ell_{\rm S}) \cos \beta_{\rm S}, \label{incl1} \\
\cos i_{\oplus} = \cos (\ell_{\oplus} - \ell_{\rm S}) \cos \beta_{\rm S}. \label{incl2}
\end{eqnarray}
The ecliptic longitude of the Sun and of the geocentric observer as seen from $T$ can be obtained from the longitude of the asteroid as seen from the barycentre of the Sun $\ell_{0 \odot}$ and of the Earth $\ell_{0 \oplus}$, respectively, given by the NASA Horizon ephemerides, as: $\ell_{\odot} = \pi - \ell_{0 \odot}$ and $\ell_{\oplus} = \pi - \ell_{0 \oplus}$. From Eqs.~(\ref{incl1}) and (\ref{incl2}), we see that both the angles $i_{\odot}$ and $ i_{\oplus}$ vary between $\beta_{\rm S}$ and $\pi - \beta_{\rm S}$ as the asteroid revolves around the Sun. 

\subsection{Radial velocity variation induced by the asteroid rotation}
\label{model}
\begin{figure}[b]
\centerline{\includegraphics[width=9cm]{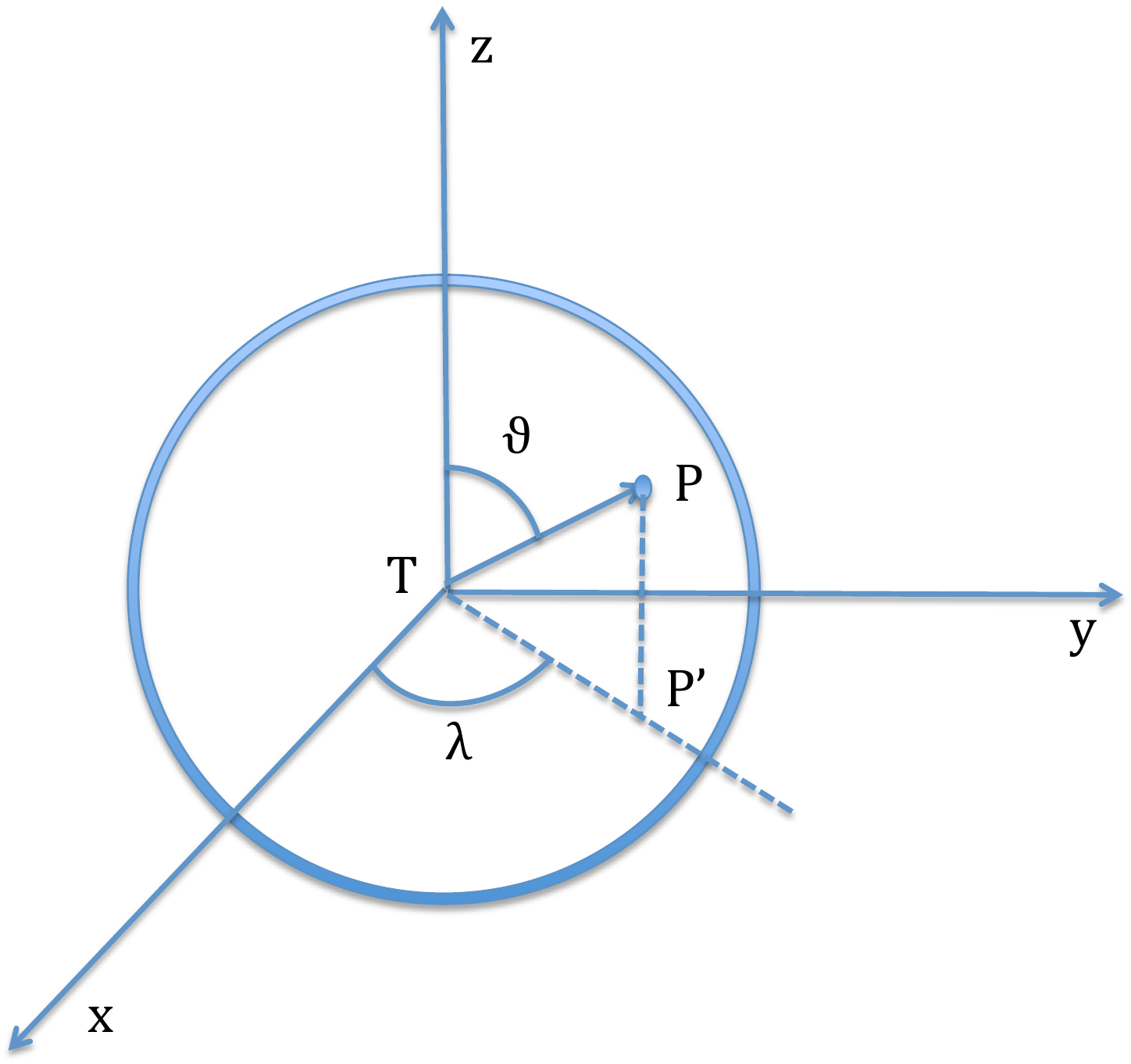}}
\vspace*{-2cm}
\caption{The adopted Cartesian reference frame with its origin at the barycentre $T$ of the asteroid, the $z$-axis along its spin axis  and the $x$-$y$ plane in its equatorial plane. The spherical coordinates of a generic point $P$ on its surface are also indicated -- $\theta$ colatitude from the North pole; $\lambda$ longitude; $P^{\prime}$ is the projection of $P$ on the equatorial plane.   }
\label{geometry}
\end{figure}

\begin{figure}[b]
\centerline{\includegraphics[width=9cm,height=12cm]{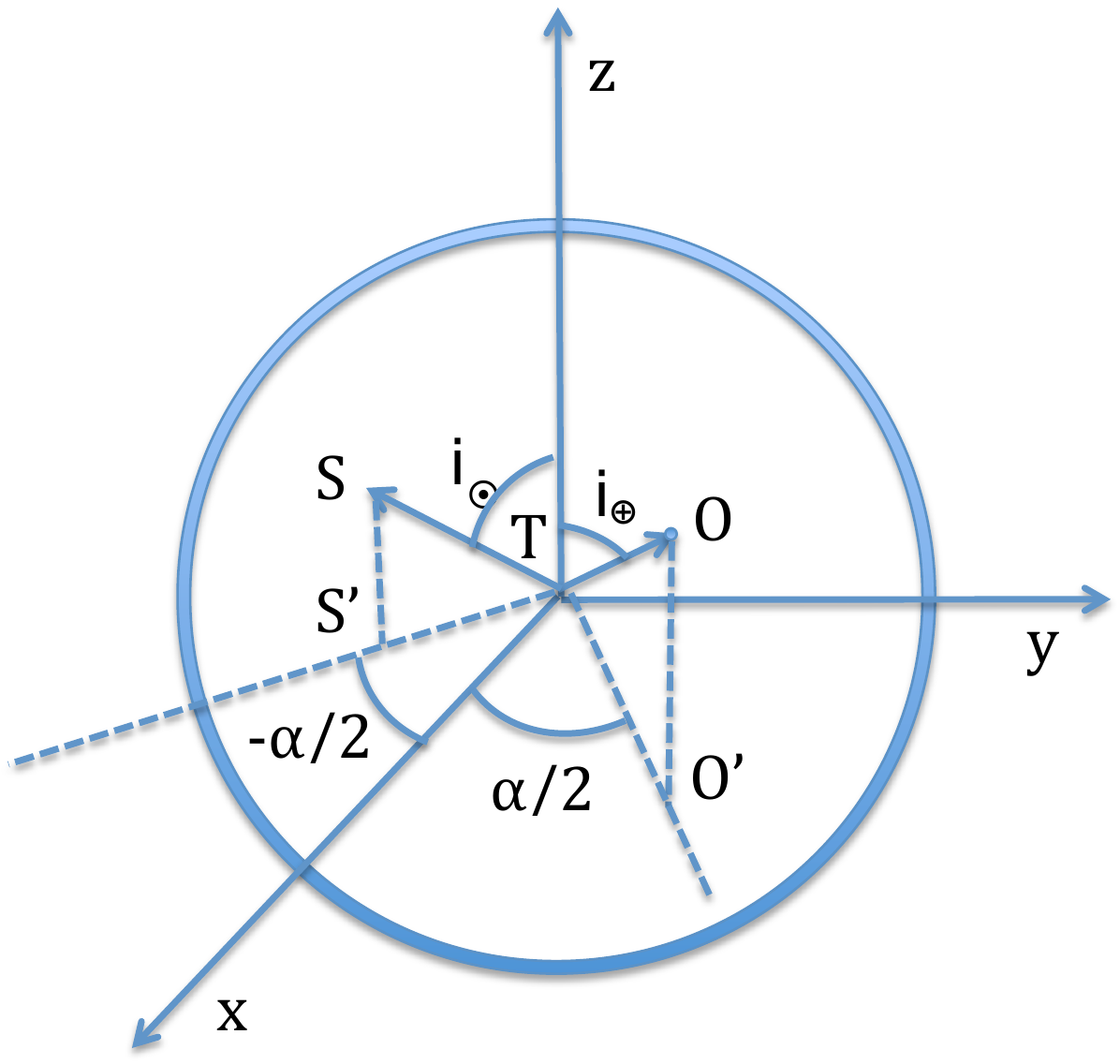}}
\vspace*{-3cm}
\caption{The sub-observer point $O$ and the sub-solar point $S$ in the adopted reference frame with the origin at the barycentre $T$ of the asteroid. The initial meridian from which the longitude is measured is chosen so that the longitude of the points $S$ and $O$ be $-\alpha/2$ and $\alpha/2$, respectively, where $\alpha$ is the angle between the directions to the observer and to the Sun with its vertex at the barycentre of the asteroid $T$ (origin of the Cartesian reference frame). The inclinations of the $TO$ and $TS$ directions to the spin axis of the asteroid are indicated as $i_{\oplus}$ and $i_{\odot}$, respectively.}
\label{sub_points}
\end{figure}

\begin{figure}[b]
\centerline{\includegraphics[width=9cm]{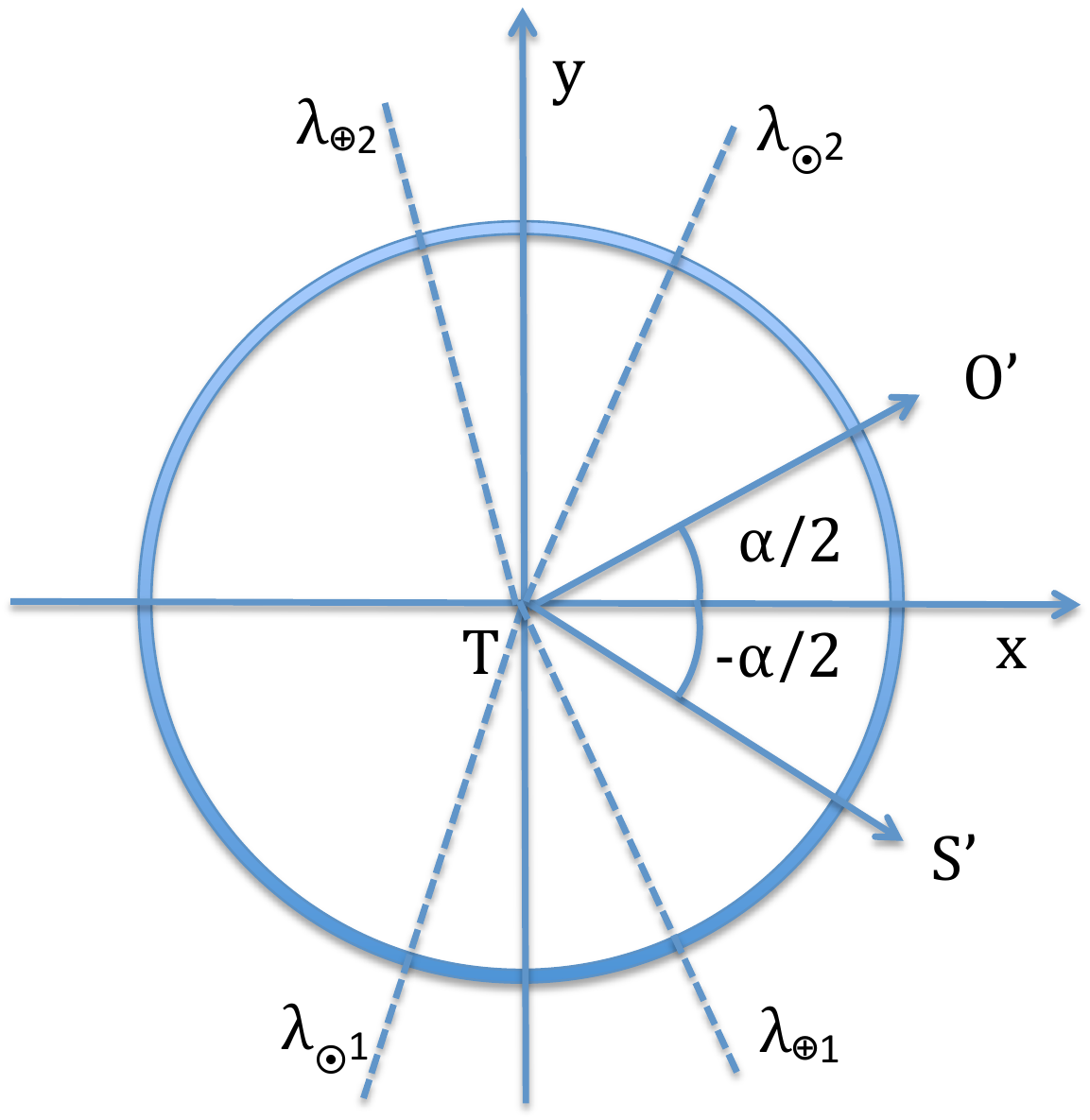}}
\vspace*{-3cm}
\caption{Projections of the $TO$ and $TS$ directions onto the equatorial plane of the asteroid ($\theta = \pi/2$).  The extreme  longitudes marking the limb of the observable disc ($\lambda_{\oplus (1,2)}$) and  of the illuminated hemisphere ($\lambda_{\odot (1, 2)}$) are indicated, respectively. The limits of integration to be considered in the RV and flux calculations are $\lambda_{1} = \max (\lambda_{\odot 1}, \lambda_{\oplus 1}) = \lambda_{\oplus 1}$ and $\lambda_{2} = \min (\lambda_{\odot 2}, \lambda_{\oplus 2}) = \lambda_{\odot 2}$, in this specific case. For $\theta \not= \pi/2$, the determination of the limits of integration is similar (see the text).  }
\label{extrema_fig}
\end{figure}

We consider a Cartesian reference frame with the origin at $T$, the $z$-axis along the spin axis of the asteroid and the $x$ and $y$ axes in the equatorial plane of the asteroid. In addition to the Cartesian coordinates, we consider also spherical coordinates  (cf. Fig.~\ref{geometry}). The origin of the longitude is chosen in such a way that the longitude of the centre of the Sun is $-\alpha/2$, while that of the observer is $\alpha/2$  at the time of  sunlight reflection (cf. Fig.~\ref{sub_points}) -- a similar approach has been introduced to model asteroid photometric variations \citep[cf., ][]{Harrisetal84}. Therefore, the unit vectors from $T$ to the centre of the Sun $S$ and to the geocentric observer $O$ have Cartesian components:
\begin{eqnarray}
\label{to}
\hat{TO} & = & (\sin i_{\oplus} \cos (\alpha/2), \sin i_{\oplus} \sin (\alpha/2), \cos i_{\oplus}), \\
\hat{TS} & = & (\sin i_{\odot} \cos (\alpha/2), -\sin i_{\odot} \sin (\alpha/2), \cos i_{\odot}).  
\label{ts}
\end{eqnarray}
A generic point on the surface of the  asteroid has Cartesian coordinates $ P \equiv R (\sin \theta \cos \lambda, \sin \theta \sin \lambda, \cos \theta)$, where  $0 \leq \theta \leq \pi$ is its colatitude measured from the North pole, $0 \leq \lambda \leq 2\pi$ its longitude, and $R$ the radius of the body assumed to be spherical (see Fig.~\ref{geometry}). 
Let us consider the angles $\psi_{\oplus}$ between the normal in $P$ and the direction to the observer $\hat{TO}$,  and $\psi_{\odot}$ between the normal and the direction to the Sun $\hat{TS}$. By performing the scalar products between the unit vectors $\hat{OP}$ and $\hat{TO}$, and $\hat{OP}$ and $\hat{TS}$, the Cartesian components of which are given above, we find:
\begin{eqnarray}
\label{mu_def1}
\mu_{\oplus} \equiv \cos \psi_{\oplus} & = & \sin i_{\oplus} \sin \theta \cos(\lambda -\alpha/2) + \cos \theta \cos i_{\oplus}, \\
\mu_{\odot} \equiv \cos \psi_{\odot} & = & \sin i_{\odot} \sin \theta \cos(\lambda + \alpha/2) + \cos \theta \cos i_{\odot}. 
\label{mu_def2}
\end{eqnarray}
The rotation velocity of the point $P$ is found by differentiating its position vector  as a function of the time because the longitude of $P$ increases steady in our fixed reference frame owing to the rotation of the asteroid. Introducing the equatorial rotation velocity $V_{\rm eq} \equiv (2\pi/P_{\rm rot}) R$, where $P_{\rm rot}$ is the rotation period, we find:
\begin{equation}
{\bf V}(P) = V_{\rm eq} (-\sin \theta \sin \lambda, \sin \theta \cos \lambda, 0). 
\end{equation}
The measured RV is a weighted average over the illuminated portion of the asteroid's disc, where the weight of each disc element is proportional to the flux received from it. Therefore, we need to define the limits of the visible disc from which a non-zero flux is received by the observer. The limb of the visible disc is defined by the condition $\mu_{\oplus} = 0$ that gives the longitude limits $\lambda_{\oplus (i)}(\theta)$ of the disc for a given colatitude $\theta$ as: 
\begin{equation}
\cos (\lambda_{\oplus (i)} - \alpha/2) = -\cot i_{\oplus} \cot \theta, \; \mbox{$i = 1, 2.$} 
\label{eqlo}
\end{equation}
On the other hand, the limb of the illuminated hemisphere of the asteroid is defined by the condition: $\mu_{\odot} = 0$. Indicating with $\lambda_{\odot (k)}$ the longitude limits of the illuminated hemisphere at a given colatitude $\theta$, we find: 
\begin{equation}
\cos (\lambda_{\odot (k)} + \alpha/2) = -\cot i_{\odot} \cot \theta, \; \mbox{$k= 1, 2.$}
\label{eqls}
\end{equation} 
We introduce the angles $\gamma_{\oplus} \equiv \arccos (-\cot i_{\oplus} \cos \theta)$ and $\gamma_{\odot} = \arccos (-\cot i_{\odot} \cot \theta)$. 
For a given colatitude $\theta$ on the disc of the asteroid, the {\it visible and illuminated} longitude range $[\lambda_{1}(\theta), \lambda_{2}(\theta)]$ is given by (cf.~Fig.~\ref{extrema_fig}, where we show the case of the equatorial plane, i.e., $\theta = \pi/2$): 
\begin{equation}
[\lambda_{1}, \lambda_{2}] = [\max(-\gamma_{\oplus}+\alpha/2, -\gamma_{\odot}-\alpha/2), \min(\gamma_{\oplus}+\alpha/2, \gamma_{\odot}-\alpha/2)].
\end{equation}
%%%%%%%%%%%%%%%%%%%%%%%%%%%%%%
%%%%%%%%%%%%%%%%%%%%%%
%%%%%%%%%%%%%%%%%%%%%%%%%%%%%%
%%%%%%%%%%%%%%%%%%%%%%
%%%%%%%%%%%%%%%%%%%%%%%%%%%%%%
%%%%%%%%%%%%%%%%%%%%%%
\noindent
~\\
A given point $P$ on the illuminated and visible disc of the asteroid receives the light coming from the Sun with a Doppler shift corresponding to its radial velocity with respect to the barycentre of the Sun, $ V_{\rm R (rot) 1}(P)$, and reflects the spectrum towards the observer without any further shift in its rest frame. For simplicity, we consider here only the component of the radial velocity produced by the rotation of the body, that we indicate with the subscript (rot). The observer on the Earth is moving with respect to the reflecting point with a radial velocity $V_{\rm R (rot)}(2)$, also due to the rotation of the body, that introduces a further Doppler shift in the spectrum. In conclusion, the observed spectrum coming from $P$ is Doppler shifted with a radial velocity corresponding to the sum of  the two contributions, i.e., $ V_{\rm R (rot) (1)}(P) + V_{\rm (rot) (2)} (P)$ \citep[cf.][]{Nieuw69,Gjurch05,Gjurch13}. 

The perturbation of the measured RV produced by the rotation of the reflecting body is  found by averaging the above Doppler shift over the surface of its visible and illuminated disc, i.e.: 
\begin{equation}
\Delta RV = \frac{\int_{D_{\rm I}} [ V_{\rm R (rot) (1)}(P) + V_{\rm (rot) (2)} (P) ] d F(P) }{\int_{D_{\rm I}} dF(P)},
\label{vr}
\end{equation}
where  $dF(P)$ is the  flux coming from the area element of the disc around the point $P$;  the integration is extended over the illuminated portion of the disc $D_{\rm I}$. 

The flux coming from a given surface element at frequency $\nu$ is given by:
\begin{equation}
dF(P) =  I(\nu, P) \mu_{\oplus}(P) dA(P),
\label{dflux}
\end{equation}
where  $I(\nu, P)$ is the specific intensity of  the element  at $\nu$ and $dA(P) = R^{2} \sin \theta \, d\theta d\lambda$ the area of the surface element. 

Since the asteroid is illuminated by the Sun, the intensity reflected at a given point of its surface is given by Lambert's law \citep[e.g.,][]{Kopal59}: 
\begin{equation}
I(\nu, P) = I_{0}(\nu) a(\nu) \cos \psi_{\odot} = I_{0}(\nu) a(\nu) \mu_{\odot},  
\label{lambert}
\end{equation}
where $I_{0}(\nu)$ is the intensity for normal reflection and $a(\nu)$ the albedo at frequency $\nu$.  
The total radial velocity difference produced at the point $P$ by the rotation of the asteroid is the sum of the projections $V_{\rm R (rot) (1)} = {\bf V}(P) \cdot \hat{TO}$ and $V_{\rm R (rot) (2)}  = {\bf V}(P) \cdot \hat{TS}$; therefore, we obtain: 
\begin{equation}
V_{\rm R (rot) (1)} + V_{\rm R (rot) (2)}  = - V_{\rm eq} \sin \theta \left[ \sin i_{\oplus} \sin (\lambda -\alpha/2) + \sin i_{\odot} \sin (\lambda + \alpha/2) \right]. 
\label{vr_tot} % vr_tot
\end{equation} 
The  integral in the numerator of Eq.~(\ref{vr}) at a given frequency $\nu$ becomes:
\begin{eqnarray}
\lefteqn{\int_{D_{\rm I}} [V_{\rm R (rot) (1)}(P) + V_{\rm (rot) (2)} ] \,  dF(P) = } \nonumber \\
& =  & -V_{\rm eq} I_{0}(\nu) a(\nu) R^{2}  \int_{0}^{\pi}  d\theta \int_{\lambda_{1}(\theta)}^{\lambda_{2}(\theta)} d\lambda \sin^{2} \theta \, \left[ \sin i_{\oplus} \sin(\lambda -\alpha/2) \right.  \\
& + & \left. \sin i_{\odot} \sin (\lambda +\alpha/2) \right] \mu_{\oplus} \mu_{\odot} \, \nonumber, 
 \label{eq4_gen}
\end{eqnarray}
where the limits of integration $\lambda_{1}$ and $\lambda_{2}$ have been specified above. In general, this integral  can be evaluated only numerically owing to the non-closed expressions giving the integration limits $\lambda_{1,2}$ as a function of $\theta$. The total flux $F$ at frequency $\nu$ received from the disc of the asteroid that appears in the denominator of Eq.~(\ref{vr}), becomes: 
\begin{equation}
F  = I_{0}(\nu) a(\nu)  R^{2} \int_{0}^{\pi} d\theta \, \int_{\lambda_{1}(\theta)}^{\lambda_{2}(\theta)} d\lambda \,  \mu_{\oplus} \mu_{\odot}  \sin \theta,
\end{equation}
that can also be integrated numerically. 

In conclusion, the radial velocity perturbation  is given by:
\begin{eqnarray}
\lefteqn{\Delta RV =} \nonumber \\
& = & -V_{\rm eq} \frac{\int_{0}^{\pi}  \int_{\lambda_{1}(\theta)}^{\lambda_{2}(\theta)} \sin^{2} \theta \, [\sin i_{\oplus} \sin(\lambda -\alpha/2) + \sin i_{\odot} \sin (\lambda +\alpha/2) ] \mu_{\oplus} \mu_{\odot} \, d\theta \, d\lambda}{\int_{0}^{\pi} \, \int_{\lambda_{1}(\theta)}^{\lambda_{2}(\theta)} \,  \mu_{\oplus} \mu_{\odot}  \sin \theta \, d\theta \, d\lambda,
}
\label{delta_rv_final}
\end{eqnarray}
that is independent of the albedo and the specific intensity at the given frequency. Therefore, this RV shift can be applied to the whole spectrum. 

\subsection{A particular case}

When the spin axis of the asteroid is orthogonal to the plane of the ecliptic, i.e., $\beta_{\rm S}=\pi/2$, Eqs.~(\ref{incl1}) and (\ref{incl2}) gives: $i_{\oplus} = i_{\odot} = \pi/2$. A more general case arises when the longitudes of the Sun and of the observer are such that $i_{\oplus} = i_{\odot} = i$, where the common inclination $i$ is arbitrary. Now, we shall consider this particular case. 

Comparing Eqs.~(\ref{eqlo}) and (\ref{eqls}), we see that: 
\begin{equation}
\lambda_{\oplus (i)} = -\lambda_{\odot (k)} \; \mbox{with $i \not= k$.}
\label{extrema}
\end{equation}
For a given colatitude $\theta$, the longitude of the illuminated portion of the visible disc ranges from $\lambda_{1}(\theta) = \lambda(\theta)_{\oplus (1)}$ to $\lambda_{2}(\theta) = \lambda(\theta)_{\odot (2)}$  when the limb of the visible disc at $\lambda(\theta)_{\oplus (1)}$ is illuminated (cf.~Fig.~\ref{extrema_fig}), or from $\lambda_{1}(\theta) = \lambda(\theta)_{\odot (1)}$ to $\lambda_{2}(\theta) = \lambda(\theta)_{\oplus (2)}$ when the limb at $\lambda(\theta)_{\oplus (2)}$ is illuminated. 
 
 In the particular case that we are considering, the integral at the numerator of Eq.~(\ref{vr}) becomes:
\begin{eqnarray}
\lefteqn{\int_{D_{\rm I}} [V_{\rm R (rot) (1)}(P) + V_{\rm (rot) (2)} ] \,  dF(P) = } \nonumber \\
 \label{eq4}
& = &  -V_{\rm eq} I_{0}(\nu) a(\nu) R^{2} \sin i  \int_{D_{\rm I}} \sin^{2} \theta \, [\sin(\lambda -\alpha/2) + \sin (\lambda +\alpha/2) ] \mu_{\oplus} \mu_{\odot}\, d\theta \, d\lambda \\
& = & -V_{\rm eq} I_{0}(\nu) a(\nu) R^{2} \sin i \int_{0}^{\pi}  d\theta \int_{\lambda_{1}(\theta)}^{\lambda_{2}(\theta)} d\lambda \sin^{2} \theta \, [\sin(\lambda -\alpha/2) + \sin (\lambda +\alpha/2) ] \mu_{\oplus} \mu_{\odot}\, \nonumber, 
\end{eqnarray}
where $\lambda_{1}$ and $\lambda_{2}$ were specified above and are equal to zero when the colatitude $\theta$ corresponds to a point outside the visible disc of the asteroid. 

The evaluation of the integral (\ref{eq4}) can be made by considering the symmetry of the integrand function with respect to a change in the sign of $\lambda$. 
Considering Eqs.~(\ref{mu_def1}) and (\ref{mu_def2}), the product $\mu_{\oplus} \mu_{\odot}$ is invariant under a change in the sign of $\lambda$. On the other hand, the same transformation changes the sign of $[ \sin(\lambda -\alpha/2) + \sin (\lambda + \alpha/2)]$ thus making the integrand in Eq.~(\ref{eq4}) antisymmetric with respect to  the transformation $\lambda \rightarrow -\lambda$. When we perform this change of variable in the integral, 
the limits of integration changes according to Eq.~(\ref{extrema}), say, $\lambda_{\oplus (1)}$ becomes $\lambda_{\odot (2)}$ and $\lambda_{\oplus (2)}$ becomes $ \lambda_{\odot(1)}$. Together with the sign change in the differential of the integration variable $d \lambda$, this results in no change in the limits of integration. 
This implies that the integral (\ref{eq4}) vanishes. In other words, by integrating the Doppler shift over the visible disc of the asteroid, we find that the net change in the solar RV is zero when $i_{\oplus} = i_{\odot}$. 

\subsection{An asteroid of ellipsoidal shape}
\label{model_elliptic}
\begin{figure}[b]
\centerline{\includegraphics[height=15cm]{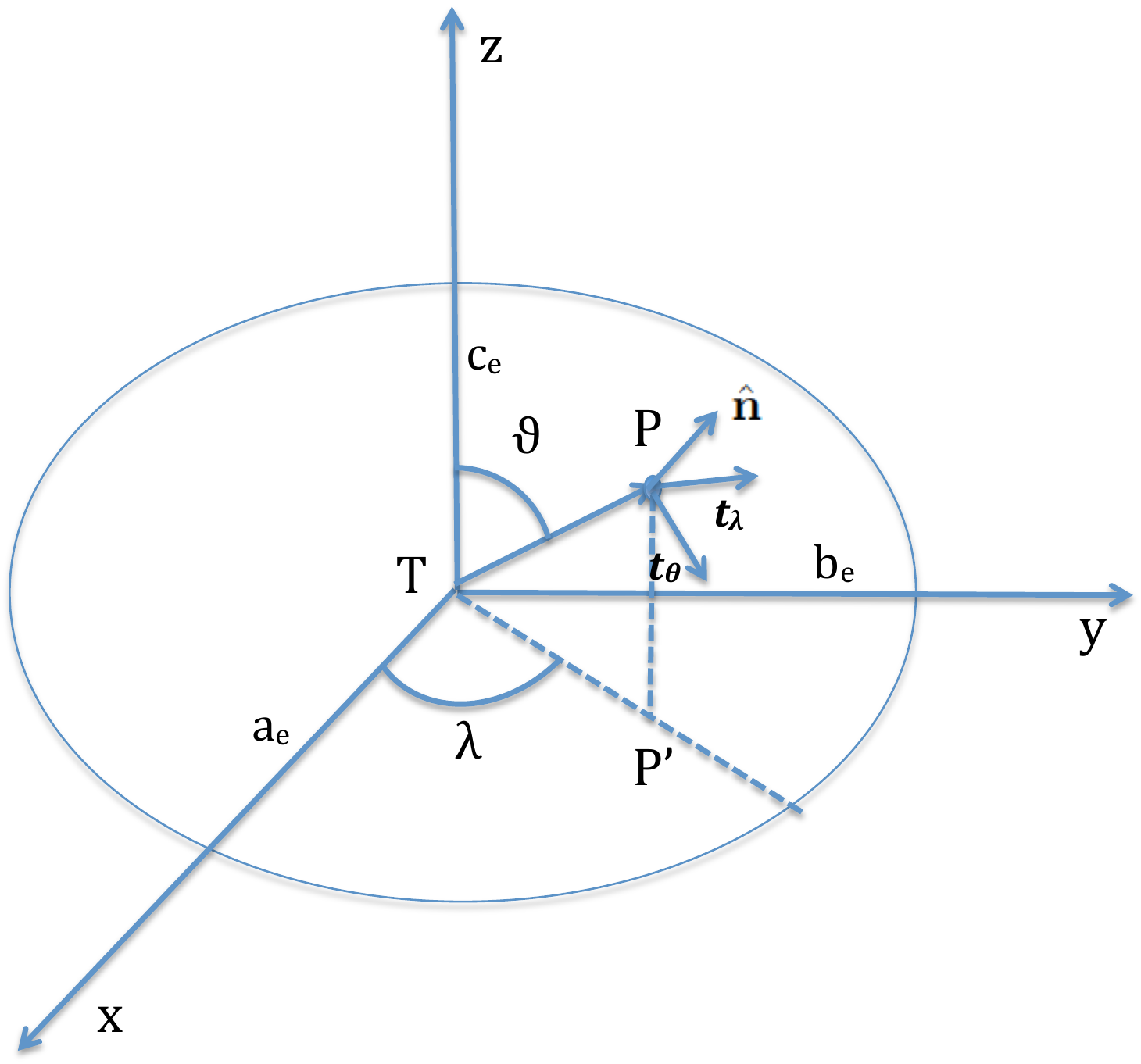}}
\vspace*{-4.5cm}
\caption{Illustration of the ellipsoidal model adopted to compute radial velocity variations in Sect.~\ref{model_elliptic}. The semiaxes of the ellipsoid $a_{\rm e}$, $b_{\rm e}$, and $c_{\rm e}$ are labelled and are assumed to be parallel to the axes of the Cartesian reference frame. The spherical polar coordinates $\lambda$ and $\theta$ of a generic point $P$ on the surface of the ellipsoid are indicated together with the local unit normal $\hat{\bf n}$ and the unit vectors ${\bf t}_{\lambda}$ and ${\bf t}_{\theta}$ tangent in $P$ in the directions of increasing longitude and colatitude, respectively. $P^{\prime}$ is the projection of $P$ onto the equatorial plane of the asteroid with its rotation axis assumed to coincide with the $\hat{z}$-axis.}
\label{ellipsoidal_geometry}
\end{figure}

The model in Sect.~\ref{model} can be extended to the case of an ellipsoidal asteroid whose  equation referred to the Cartesian reference frame with origin at $T$ is:
\begin{equation}
f(x,y,z) \equiv \frac{x^{2}}{a_{\rm e}^{2}} + \frac{y^{2}}{b_{\rm e}^{2}} + \frac{z^{2}}{c_{\rm e}^{2}} = 1,
\end{equation}
where $x,y,z$ are the Cartesian coordinates of a point $P$ on the surface of the body and $a_{\rm e}$, $b_{\rm e}$, and $c_{\rm e}$ are its semiaxes { (cf. Fig.~\ref{ellipsoidal_geometry})}. In the adopted spherical polar reference frame, the Cartesian coordinates of $P$ can be expressed as a function of its spherical coordinates as: $P\equiv (a_{\rm e} \sin \theta \cos \lambda, b_{\rm e} \sin \theta \sin \lambda, c_{\rm e} \cos \theta)$.
%%%%%%%%%%%%%%%%%%%%%%%%%%%%%%
%%%%%%%%%%%%%%%%%%%%%%
Using these parametric equations for the ellipsoidal surface, we can compute the Cartesian components of the elementary vectors ${\bf t}_{\theta}$ and ${\bf t}_{\lambda}$ that are tangent to the surface at the point $P$ in the colatitude and in the longitude directions, respectively. By differentiating the equations, we find: 
\begin{eqnarray}
{\bf t}_{\theta} & = & (a_{\rm e} \cos \lambda \cos \theta, b_{\rm e} \sin \lambda \cos \theta, -c_{\rm e} \sin \theta) d\theta \mbox{ and} \\
 {\bf t}_{\lambda} & = & (-a_{\rm e} \sin \lambda \sin \theta, b_{\rm e} \cos \lambda \sin \theta, 0 ) d\lambda.
 \end{eqnarray}
In general, ${\bf t}_{\lambda}$ and ${\bf t}_{\theta}$ are not perpendicular to each other, except when $a_{\rm e}=b_{\rm e}$. The elementary surface area at the point $P$ can be obtained as the modulus of the cross product of the two tangent vectors, i.e., $d A = | {\bf t}_{\theta} \times {\bf t}_{\lambda}|$ that can be easily computed from their Cartesian components.  

The unit normal  $\hat{\vec n}$ at $P$ is parallel to $\nabla f (x,y,z)$.  Expressing the Cartesian components of the gradient by means of the spherical coordinates: 
\begin{equation}
\hat{\bf n}(P) = \frac{1}{G}\left( \frac{\sin \theta \cos \lambda}{a_{\rm e}}, \frac{\sin \theta \sin \lambda}{b_{\rm e}}, \frac{\cos \theta}{c_{\rm e}} \right),
\end{equation}
where 
\begin{equation}
G = \sqrt{\left( \frac{\cos^{2}\lambda}{a_{\rm e}^{2}} + \frac{\sin^{2}\lambda}{b_{\rm e}^{2}} \right) \sin^{2} \theta + \frac{\cos^{2}\theta}{c_{\rm e}^{2}}}.  
\end{equation}
The expressions of the unit vectors $\hat{TO}$ and $\hat{TS}$ do not change for an ellipsoidal asteroid (see Eqs.~\ref{to} and \ref{ts}), while the projection factors $\mu_{\oplus} = \hat{\bf n} \cdot \hat{TO}$ and $\mu_{\odot} = \hat{\bf n} \cdot \hat{TS}$ can be obtained from the above components of the unit normal $\hat{\vec n}$ at each surface point. 

The components of the rotation velocity of a given point $P (\lambda, \theta)$ on the surface of the ellipsoid can be obtained by differentiating its longitude with respect to the time, as in the case of a spherical body. We obtain:
\begin{equation}
{\bf V} (P) = \frac{2\pi}{P_{\rm rot}} \left( -a_{\rm e} \sin \lambda  \sin \theta, b_{\rm e} \cos \lambda \sin \theta, 0 \right),  
\end{equation}
where $P_{\rm rot}$ is its rotation period.  This equation can be used to compute the components of the rotation velocity that appears in Eq.~(\ref{vr}) by performing the scalar products as in Sect.~\ref{model}. 

The integrations in Eq.~(\ref{vr}) can be performed numerically by dividing the ellipsoid into many surface elements and computing the contribution of each element to the radial velocity perturbation, i.e., to its numerator, and to the total flux, i.e, to its denominator. The expression for the elementary flux is not changed and is given by Eqs.~(\ref{dflux}) and (\ref{lambert}). Finally, the integrals are obtained by summing the contributions of all the surface elements that are both illuminated by the Sun ($\mu_{\odot} \geq 0$) and visible from the observer ($\mu_{\oplus} \geq 0$).
 
%\end{document}
\section{Results}
\label{applications}

For some bright asteroids, we list the mean radius, perihelion distance from the Sun, rotation period, equatorial rotation velocity,  ecliptic longitude and latitude of the spin pole with respect to the J2000 reference frame, and the corresponding reference in Table~\ref{table1}. Some of the rotation periods are taken from the NASA Horizon Ephemerides.  For 20 Massalia there is an ambiguity in the longitude of the pole $\lambda_{\rm S}$, so both values are listed. In our case, the maximum value of $V_{\rm eq} = 92.27$~m/s is achieved in the case of 1~Ceres. 

The angle $\alpha$ between the direction of the observer and that of the Sun at the time of light reflection is virtually identical to the Sun-Target-Observer angle (STO) as given by the Horizon ephemerides. 
The maximum value of   $\alpha$ is estimated as $\alpha_{\rm max} \sim \arcsin (1.016/q)$, where $q$ is the minimum heliocentric distance of the asteroid in AU. This configuration  corresponds to the Earth and the Sun seen in quadrature from the asteroid, while it is at the perihelion and the Earth at the aphelion. Using the data listed in Table~\ref{table1}, we find  $\alpha_{\rm max} \sim 31^{\circ}$. The  maximum values of $\Delta RV$ are listed in Table~\ref{max_rv} in the case of a spherical body. They were obtained by looking for the maximum radial velocity variation as given by Eq.~(\ref{delta_rv_final}) when the difference $\ell_{\odot}-\ell_{\rm S}$ is varied from $0^{\circ}$ to $360^{\circ}$ and the difference $\ell_{\oplus}-\ell_{\rm S} = ( \ell_{\odot}-\ell_{\rm S}) \pm \alpha_{\rm max}$ because these two extreme values lead to the largest radial velocity perturbation. 

\begin{figure}[b]
\centerline{\includegraphics[height=15cm]{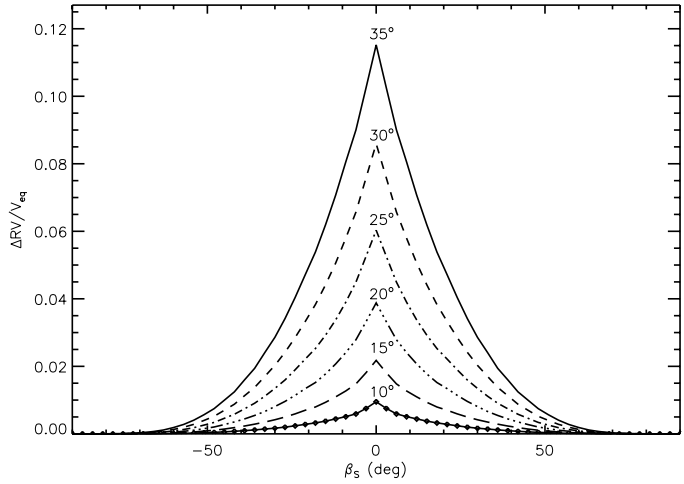}}
\vspace*{-6cm}
\caption{Maximum values of the ratio $\Delta RV/V_{\rm eq}$ vs. the ecliptic latitude $\beta_{\rm S}$ of the spin pole for a spherical asteroid. Different linestyles indicate different values of the angle $\alpha$ as labelled.  }
\label{maximum_vr-veq}
\end{figure}

The  ratio of the maximum RV perturbation to the equatorial velocity vs. the latitude of the spin pole is plotted in Fig.~\ref{maximum_vr-veq} for different values of the angle $\alpha$, considering a spherical body. This is an adequate assumption for large, bright asteroids that deviate from a spherical shape by a negligible amount for our purpose. For instance, in the case of 1~Ceres, the surface has the shape of an oblate spheroid with semiaxes $a_{\rm e} = b_{\rm e} =967 \pm 10$~km and $c_{\rm e} = 892 \pm 10$~km, i.e. an oblateness less than 10 percent \citep{Drummondetal14}. For 4~Vesta, a reference spheroid with semiaxes $a_{\rm e} = b_{\rm e} \sim 285 $~km and $c_{\rm e} \sim 229$~km provides an approximated description of the surface \citep{Jaumannetal13}. These deviations from a spherical shape produce a difference of a few cm/s at most  owing to the rather high inclination of the spin axes of 1~Ceres and 4~Vesta to the plane of the ecliptic  (cf. Table~\ref{max_rv}). In the case of 2~Pallas, the approximating ellipsoid has semiaxes: $a_{\rm e} = 275 \pm 4$, $b_{\rm e} = 258 \pm 3$, and $c_{\rm e} = 238 \pm 3$~km \citep{Carryetal10}. Together with the low inclination of its spin axis to the plane of the ecliptic, this may produce a larger deviation from the results computed with a spherical shape, of the order of a few tens of cm/s.

Other reflecting bodies often used to measure the solar RV are the Galilean satellites. 
 Their parameters are listed in Table~\ref{table2}.  Their spin axes are  almost aligned with their orbital angular momenta -- the maximum deviation is $0.46$ degrees for Europa \citep[e.g., ][]{HenrardSchwanen04} -- and their rotation periods are  synchronized with their orbital periods owing to the strong tidal interaction with Jupiter. Since their orbits are in the equatorial plane of Jupiter that is inclined by less than $\sim 4^{\circ} \, 30^{\prime}$ to the ecliptic, their spin axes are almost orthogonal to the ecliptic plane and $i_{\oplus} \sim i_{\odot} \sim 90^{\circ}$. 
Moreover,  their distance from the Sun is larger than in the case of the main-belt asteroids, thus $\alpha_{\rm max} \sim 14^{\circ}$. 
As a consequence, their maximum radial velocity variation is found to be of a few cm/s. 
%%%%%%%%%%%%%%%%%%%%%%%%%%%%%%
%%%%%%%%%%%%%%%%%%%%%%

%%%%%%%%%%%%%%%%%%%%%%%%%%%%%%%%%
\begin{table*}[t]
\begin{tabular}{rccccccc}
\hline
 & & & \\
Asteroid & q  & $R$ & $P_{\rm rot}$ & $V_{\rm eq}$ & $\ell_{\rm S}$ & $\beta_{\rm S}$ & Reference\\
& & & & & & \\
 & (AU) &  (km) & (hr) & (m/s) & (deg) & (deg)  & \\
 & & & \\
 \hline
 & & & \\
 1 Ceres & 2.55 & 480 & 9.075 & 92.27 & 346 & 82 & 1\\
 2 Pallas & 2.12 & 275 & 7.811 & 63.65 & $30$ & $-16$& 2 \\
 3 Juno & 1.98 & 117 & 7.210 & 28.31 & 108 & 38 & 3\\
 4 Vesta & 2.14 & 265 & 5.342 & 86.53 & 336 & 63 & 4\\
 7 Iris & 1.84 & 100 & 7.139 & 24.44 & 15 & 25 & 5\\
% 14  Irene & 2.16 & 76 & 15.028 &  8.82 & $-$ & $-$ & $-$ \\
 % 18  Melpomene & 1.80 & 70 & 11.570 &  10.55 & $-$ & $-$ & $-$ \\
 20 Massalia & 2.06 & 73 & 8.098 & 15.72 & 31/208 & 69 & 3\\
 & & & \\
\hline
\end{tabular}
\caption{Asteroid parameters. References: 1: \cite{Drummondetal14}; 2: \cite{Carryetal10};  3: \cite{Dottoetal95};  4: \cite{Thomasetal97}; 5: \cite{dePateretal94}.}
\label{table1}
\end{table*}
%%%%%%%%%%%%%%%%%%%%%%%%%%%%%%%%%%%%%%%
%%%%%%%%%%%%%%%%%%%%%%%%%%%%%%%%%
\begin{table}
\begin{tabular}{ccc}
\hline
 & &  \\
Asteroid & $\alpha_{\rm max}$ & $\Delta RV_{\rm max}$\\
& & \\
 & (deg) & (m/s) \\
 & &  \\
 \hline
 & &  \\
 1 Ceres & 23.5 & $1.31 \times 10^{-4}$\\
 2 Pallas & 28.6 & 2.428 \\
 3 Juno & 30.9 & 0.361\\
 4 Vesta & 28.3 & 0.060\\
 7 Iris & 28.9 & 0.602 \\
 20 Massalia & 29.6 & $3.89 \times 10^{-3}$ \\
 & &  \\
\hline
\end{tabular}
\caption{Maximum RV perturbation for the asteroids in Table~\ref{table1}. }
\label{max_rv}
\end{table}
%%%%%%%%%%%%%%%%%%%%%%%%%%%%%%%%%%%%%%%

%%%%%%%%%%%%%%%%%%%%%%%%%%%%%%%%%
\begin{table}
\begin{tabular}{cccc}
\hline
 & & & \\
Satellite & $R$ & $P_{\rm rot}$ & $V_{\rm eq}$ \\
 & (km) & (day) & (m/s) \\
 & & & \\
 \hline
 & & & \\
Io & 1821 & 1.77 & 74.84 \\
Europa & 1605 & 3.55 & 31.98 \\
Ganimede & 2631 & 7.16 & 26.72 \\
Callisto & 2411 &  16.69 & 10.50 \\
 & & & \\
\hline
\end{tabular}
\caption{Galileian satellite parameters. }
\label{table2}
\end{table}
%%%%%%%%%%%%%%%%%%%%%%%%%%%%%%%%%%%%%%%
\section{Discussion and conclusions}
We provide a method to compute the radial velocity perturbation induced by the rotation of a solar system body  on the reflected solar spectrum. We focus on the case of a body with a uniform albedo and find that the perturbation depends in a complex way on the direction of its spin axis and the angle Sun-body-Earth. We treat both the case of a spherical reflecting body and of a triaxial ellipsoid. The introduced approach is based on vector scalar products rather than on the application of spherical trigonometry. With a suitable choice of the reference frame, we make the  Cartesian components of the relevant vectors very simple so that the scalar products are easily computed. 

In the particular case when the angles $i_{\odot}$ and $i_{\oplus}$ of the body spin axis to the directions of the Sun and the observer, respectively, are equal, we show that the radial velocity perturbation vanishes. On the other hand, when  the inclination of the body spin axis to the plane of the ecliptic is less than about $\pm 40^{\circ}$ and the Sun-body-Earth angle $\alpha \geq 30^{\circ}$, the radial velocity perturbation can reach up to $\sim 0.1$ of the equatorial rotation velocity of the body (cf. Fig.~\ref{maximum_vr-veq}). In the case of asteroid 2~Pallas that has a spin inclination of $\sim 30^{\circ}$, the perturbation can reach $\sim 2.4$~m/s. For other bright asteroids, such as 1~Ceres and 4~Vesta, the effect is remarkably smaller, thanks to the rather high inclination of their spin axes to the plane of the ecliptic (cf. Table~\ref{max_rv}). However, in the case of 4~Vesta, it is necessary to take into account the present effect when an accuracy of a few cm/s is required. 
An easy estimate of the amplitude of the effect of rotation can be obtained from Fig.~\ref{maximum_vr-veq} when the inclination $\beta_{\rm S}$ of the spin axis to the ecliptic plane and the maximum value of the Sun-body-Earth angle $\alpha_{\rm max}$ are known (cf. Sect.~\ref{applications}). 

\begin{acknowledgements}
The authors are grateful to an anonymous referee for useful comments on their work. 
AFL gratefully acknowledges support by the Italian National Institute for Astrophysics (INAF) through the {\it Progetti premiali} funding scheme of the Italian Ministry of Education, University and Research. 

\end{acknowledgements}

\end{document}